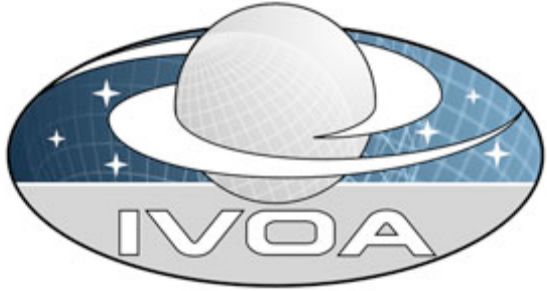

*International*
*Virtual*
*Observatory*
*Alliance*

# Simple Spectral Lines Data Model

# Version 1.0

*Recommendation 2 Dec 2010*

**This version:**
    REC-SSLDM-1.0-20101202
**Latest version:**
    http://www.ivoa.net/Documents/SSLDM
**Previous versions:**

**Editors:** Pedro Osuna, Jesus Salgado

**Authors:**
    Pedro Osuna
    Matteo Guainazzi
    Jesus Salgado
    Marie-Lise Dubernet
    Evelyne Roueff

## Status of This Document

This is an IVOA Recommendation. It has been endorsed by the IVOA Executive Committee as an IVOA Recommendation. It is a stable document and may be used as reference material or cited as a normative reference from another document. IVOA's role in making the Recommendation is to draw attention to the specification and to promote its widespread deployment. This enhances the functionality and interoperability inside the Astronomical Community.It is appropriate to reference this document only as a recommended standard that is under review and which may be changed before it is accepted as a full recommendation.

# Abstract

This document presents a Data Model to describe Spectral Line Transitions in the context of the Simple Line Access Protocol defined by the IVOA (c.f. Ref[13] IVOA Simple Line Access protocol)

The main objective of the model is to integrate with and support the Simple Line Access Protocol, with which it forms a compact unit. This integration allows seamless access to Spectral Line Transitions available worldwide in the VO context.

This model does not provide a complete description of Atomic and Molecular Physics, which scope is outside of this document.

In the astrophysical sense, a ***line*** is considered as the result of a ***transition*** between two energy ***levels***. Under the basis of this assumption, a whole set of objects and attributes have been derived to define properly the necessary information to describe lines appearing in astrophysical contexts.

The document has been written taking into account available information from many different Line data providers (see acknowledgments section).

# Acknowledgments

The authors wish to acknowledge all the people and institutes, atomic and molecular database experts and physicists who have collaborated through different discussions to the building up of the concepts described in this document.

# Contents

# 1   Introduction

Atomic and molecular line databases are a fundamental component in our process of understanding the physical nature of astrophysical plasmas. Density, temperature, pressure, ionization state and mechanism, can be derived by comparing the properties (energy, profile, intensity) of emission and absorption lines observed in astronomical sources with atomic and molecular physics data.
The latter have been consolidated through experiments in Earth's laboratories, whose results populate a rich wealth of databases around the world. Accessing the information of these databases in the Virtual Observatory (VO) framework is a fundamental part of the VO mission.

This document aims at providing a simple framework, both for atomic and molecular line databases, as well as for databases of observed lines in all energy ranges, or for VO-tools, which can extract emission/absorption line information from observed spectra or narrow-band filter photometry.

The Model is organized around the concept of "**Line**", defined as the results of a **transition** between two **levels** (this concept applies to bound-bound and free-bound transitions, but not free-free transitions). In turn each "Level" is characterized by one (or more) "**QuantumState**". The latter is characterized by a proper set of "**QuantumNumber**".

The object "**Species**" represents a placeholder for a whole new model to represent the atomic and molecular properties of matter. This will take form in a separate document. We reserve here one single attribute for the time being, the name of the species (including standard naming convention for ionised species), and shall be pointing to the future model whenever available.

Any process which modifies the intrinsic properties of a "Line" (monochromatic character, laboratory wavelength etc.) is described through the attributes of "**Process**", which allows as well to describe the nature of the process responsible for the line generation, whenever pertinent. The element "**Environment**" allows service providers to list physical properties of the line-emitting/absorbing plasma, derived from the properties of the line emission/absorption complex. Both "Process" and "Environment" contain hooks to VO "**Model**"s for theoretical physics (placeholders for future models).

The present Simple Spectral Line Data Model does not explicitly address non-electromagnetic transitions.

***Note on Quantum Numbers***: *throughout the text, Quantum Numbers appear in some formulae; their definitions can be found in Chapter 4. In particular, for the most commonly used n-electron system quantum numbers, S, L and J represent the* **total spin (S)** *(see* **4.2.1***),* **total electronic orbital momentum (L)** *(see* **4.2.4***) and* **total angular momentum (J)** *(see* **4.2.10***)*.

As already stated, this data model does not provide a complete description of Atomic and Molecular Physics, but makes reference to it, pending its completion. The following image shows the overal situation of this data model within the overall IVOA Architecture:

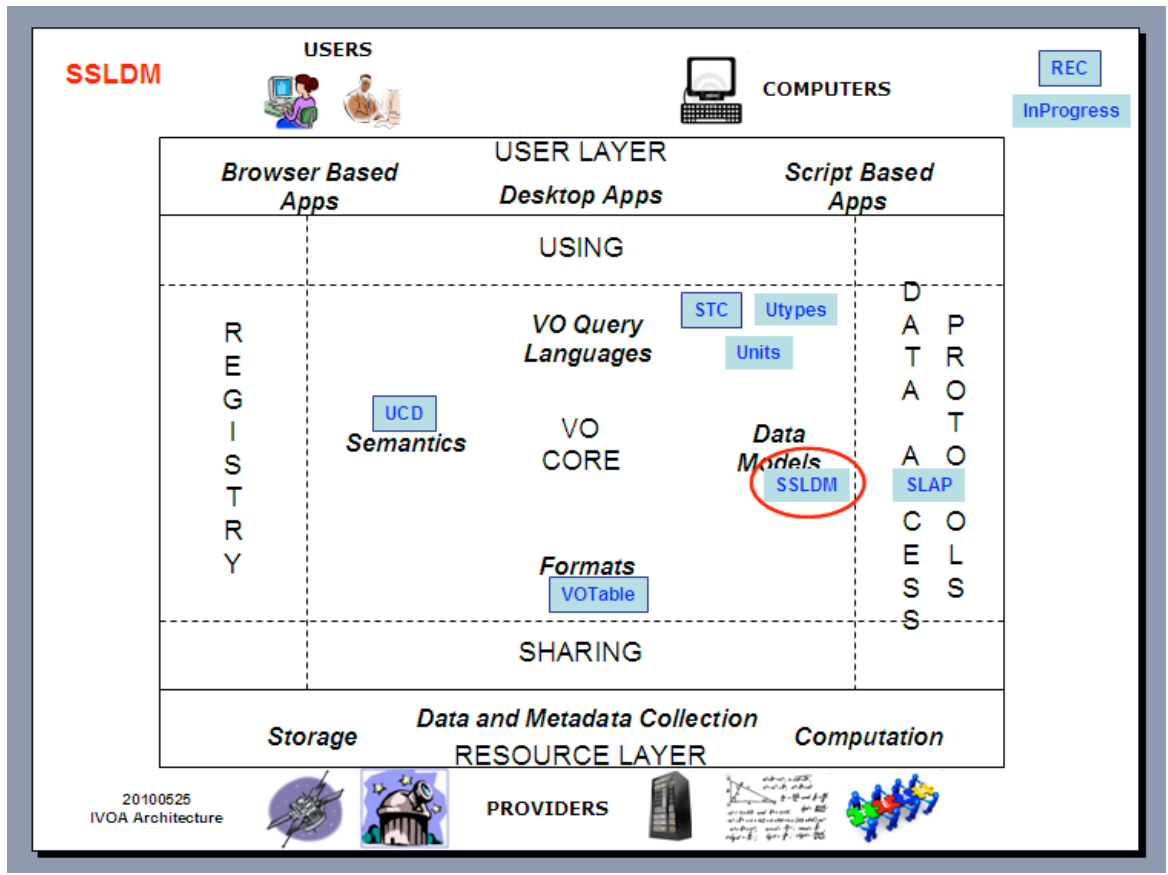

As most of the VO Data Models, SpectrumDM makes use of STC, Utypes, Units and UCDs. SSLDM can be serialized with a VOTable

# 2 Graphical representation of the Data Model

A summary of the conceptual data model for the simple spectral lines follows. Detailed descriptions of each of the attributes will be given in subsequent paragraphs.

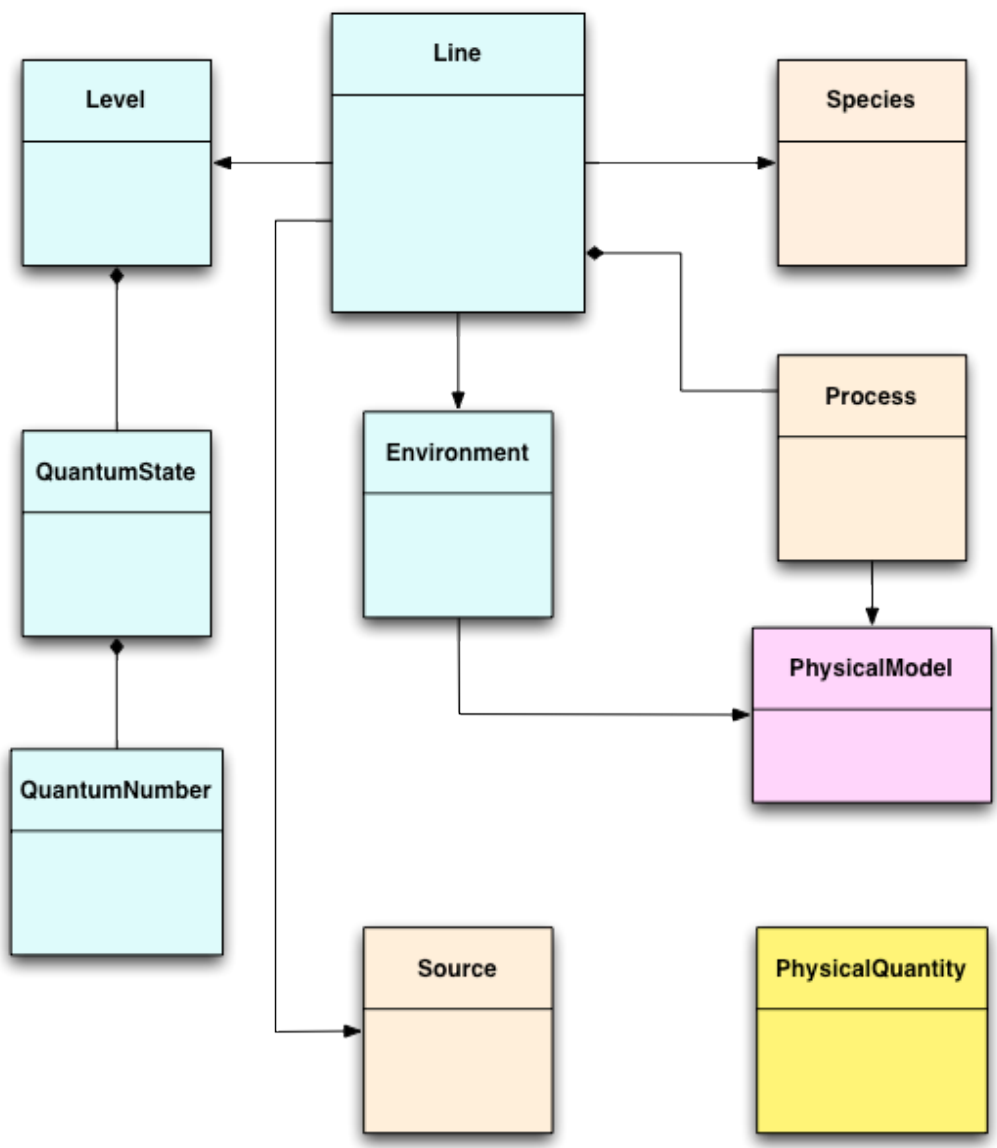

The colors in this diagram represent the different different qualification of the classes within the model:

- **Light blue** represent **main classes** within the model. They are the most important to describe properly a line within this data model. As many as possible of its attributes (see below for further details) should be filled in
- **light orange** represent **support classes**, which eventually can point to further data models to be developed by the IVOA. They have a minimum amount of attributes that is enough for this simple line data model, but can make use of possible future developments of the models they point to.
- **light pink** represent **placeholder classes**, and have no attributes in the current model (they are left here as reference for future possible developments in the area).
- **yellow** represents a **special class** that supports any physical quantity within the model. This class has not yet been developed in full by the IVOA but a self-contained definition is given here so the model can be effectively used (see details below for this class)

A detailed UML with the corresponding attributes follows:

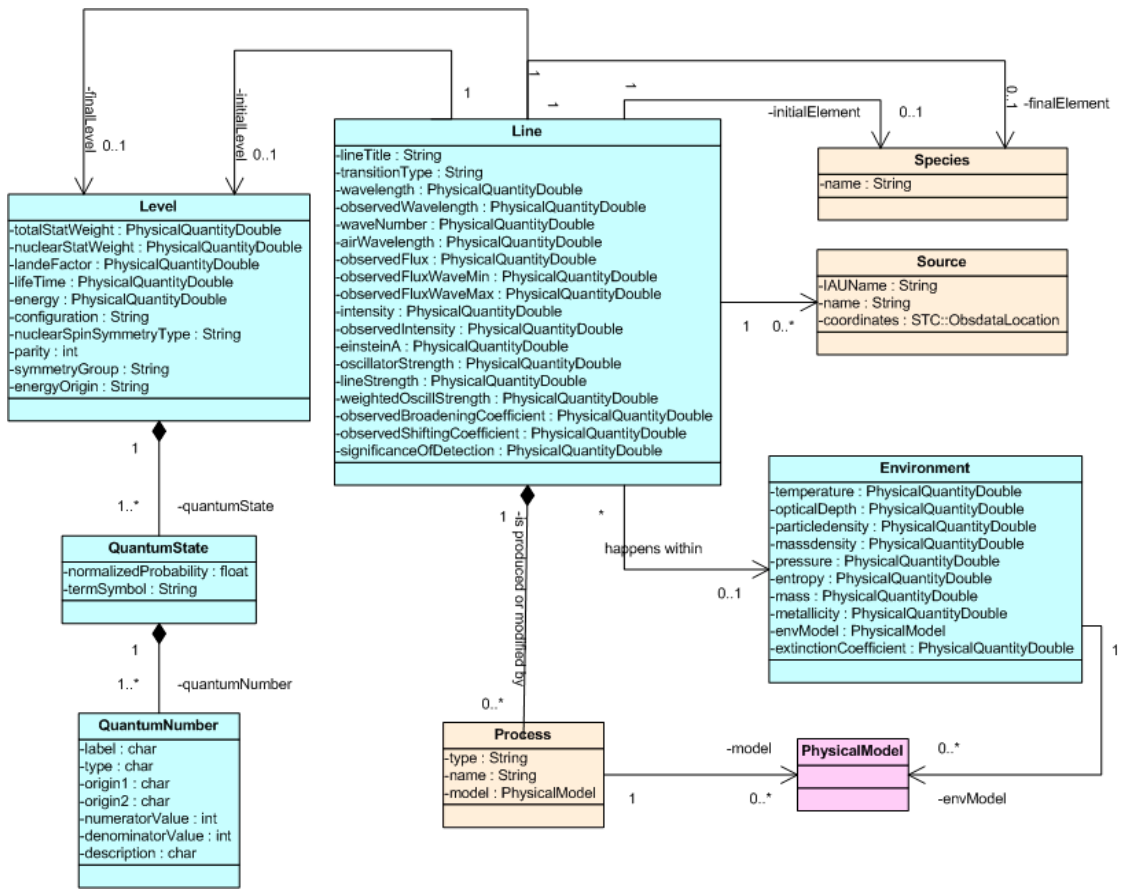

# 3 Detailed description of data model classes and attributes

## 3.1 *PhysicalQuantity*

Class used to describe a physical measurement. This could be superseded by an eventual IVOA Quantity DM definition. It contains the basic information to understand a Quantity.

The UML for Physical Quantity follows

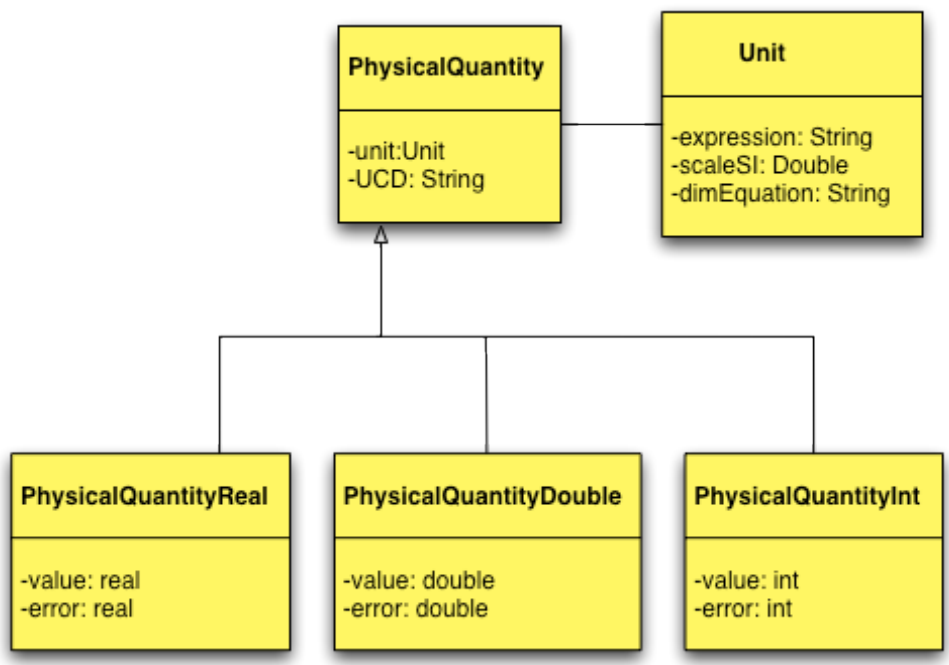

### 3.1.1 PhysicalQuantity.value: Real/Double/Integer

Value of the measure. Type of value will be:

- Real for PhysicalQuantityReal
- Double for PhysicalQuantityDouble
- Integer for PhysicalQuantityInt

Example: `2.32`

### 3.1.2 PhysicalQuantity.error: Real/Double/Integer

General error of the measure. Please note that this is the total error. A more formal description should be provided in a general IVOA Physical Quantity Data Model. Type of value same as that of the PhysicalQuantity.

Example: `0.01`

### 3.1.3 PhysicalQuantity.unit: Unit

Unit in which the measure is expressed. See the **Unit** class for definition and attributes. Both value and error should be expressed in the same units

## 3.2 Unit

Class used to describe a physical unit. This could be superseded by a general IVOA Unit DM definition. It contains the basic information to understand a Unit.

### 3.2.1 Unit.expression: String

String representation of the unit.

Example: "Jansky"

### 3.2.2 Unit.scaleSI: Double

Scaling factor to convert the unit to its International System (IS) of Units equivalent.

For instance, 1 cm has a scale factor of E-2, since 1cm = 0.01 m (its IS equivalent)..

Example: 1.`E-2`

### 3.2.3 Unit.dimEquation: String

Dimensional equation representation of the unit. The format is a string with the dimensional equation, where M is mass, L is length, T is time, K is temperature. For ease of notation, the caret "^" indicating powers of ten can be removed -as is customary in Dimensional Analysis practises- resulting in expressions like the following, which are equivalent:

Example: "`ML-1T-3`"
Example: "`ML^-1T^-3`"

> Examples of PhysicalQuantity instances:
>
> Wavelength: 4830 ± 10 Angstrom
>
> Units:
> $1 Angstrom = 10^{-10} m$
> **1.E-10 L**
>
> PhysicalQuantity.value = 4830
> PhysicalQuantity.error = 10
> PhysicalQuantity.unit.expression= Angstrom
> PhysicalQuantity.unit.scaleSI=1.E-10
> PhysicalQuantity.unit.dimEquation=L
>
> Flux Φ(ν): 3500±14 Jansky
>
> Units:
> $1 Jansky = 10^{-26} W m^{-2} Hz^{-1} = 10^{-26} Kg.s^{-2}$
> **1.E-26 MT-2**
>
> PhysicalQuantity.value = 3500
> PhysicalQuantity.error = 14
> PhysicalQuantity.unit.expression= Jansky
> PhysicalQuantity.unit.scaleSI=1.E-26
> PhysicalQuantity.unit.dimEquation=MT-2

(See, e.g., IVOA SSAP for more examples)

## 3.3 Line

This class includes observables, e.g. measured physical parameters, describing the line, as well as the main physical properties of the transition originating it. Recombination and dissociations are expressed through atomic coefficients rather than through global properties.

We define a Line to mean "the radiation associated with a possible transition between the states belonging to two levels"

### 3.3.1 Line.title: String

A small description title identifying the line. This is useful when identification is not secure or not yet established.

Example: "`Hyperfine N2H+ J1->0 transition line`"

### 3.3.2 Line.initialLevel: Level

A full description of the initial level of the transition originating the line. See the **Level** class for a definition of the Level concept and attributes.

### 3.3.3 Line.finalLevel: Level

A full description of the final level of the transition originating the line. See the **Level** class for a definition of the Level concept and attributes.

### 3.3.4 Line.initialElement: Species

A full description of the initial state of the atom (including its ionization state) or molecule, where the line transition occurs.

Example: "`N2H+`"

### 3.3.5 Line.finalElement: Species

A full description of the final state of the atom (including its ionization state) or molecule, where the line transition occurs. For bound-bound atomic transitions, it follows: "initialElement"="finalElement".

Example: "N2H+"

### 3.3.6 Line.wavelength: PhysicalQuantity
Wavelength in the vacuum of the transition originating the line.

### 3.3.7 Line.frequency: PhysicalQuantity
Frequency in the vacuum of the transition originating the line.

### 3.3.8 Line.wavenumber: PhysicalQuantity
Wavenumber in the vacuum of the transition originating the line.

### 3.3.9 Line.airWavelength: PhysicalQuantity
Wavelength in the air of the transition originating the line.

### 3.3.10 Line.einsteinA: PhysicalQuantity
Einstein coefficient $A_{ji}$ for spontaneous emission from level j (`finalLevel`) to level i (`initialLevel`), defined as the probability per second that an atom in level j will decay to level i. The Einstein A coefficient can be determined from the wave functions $\Psi_i$ and $\Psi_j$ of the initial and final states as follows:

$$A_{ji} = \frac{64\pi^4 e^2}{3hc^3 4\pi\varepsilon_0} v^3 \left| \int \Psi_j^* \vec{r} \Psi_i \, dV \right|^2$$

where $\vec{r}$ is the position vector and the integral is over the volume, thus:

$$A_{ji} = \frac{64\pi^4 e^2}{3hc^3 4\pi\varepsilon_0} v^3 |d^2| = 1.046 \times 10^{21} v^3 |d^2|$$

where:

$$d = e \int \Psi_j^* \vec{r} \Psi_i \, dV$$

is the *dipole moment*.

### 3.3.11 Line.oscillatorStrength: PhysicalQuantity

In ultraviolet and optical stellar astronomy, the *oscillator strength* $f_{ji}$ appears for purely historical reasons. In pre quantum-mechanical physics, an electromagnetic wave was described as an electron vibrating like a harmonic oscillator with energy being absorbed through damping; the fraction of energy absorbed was fixed irrespective of the transition, so an oscillator strength was introduced to allow for the fact that experimental lines have different strengths.

The *oscillator strength* is defined as:

$$f_{ji} = \frac{4\pi\varepsilon_0 mc^3}{8\pi^2 e^2} \frac{1}{v^2} \frac{g_j}{g_i} A_{ji} = 1.347 \times 10^{21} \frac{1}{v^2} \frac{g_j}{g_i} A_{ji}$$

where $g_i$ and $g_j$ represent the statistical weights of the lower and upper levels.

### 3.3.12 Line.weightedOscillatorStrength: PhysicalQuantity

The product between `oscillatorStrength` and the statistical weight $g$ of the `initialLevel`

### 3.3.13 Line.intensity: PhysicalQuantity

This is a source dependent relative intensity, useful as a guideline for low density sources. These are values that are intended to represent the strengths of the lines of a spectrum as they would appear in emission. They may have been normalized. They can be expressed in absolute physical units or in relative units with respect to a reference line.

The difficulty of obtaining reliable relative intensities can be understood from the fact that in optically thin plasmas, the intensity of a spectral line is proportional to:

$$I_{ik} = N_k A_{kl} h\nu_{ik}$$

Where $N_k$ is the number of atoms in the upper level $k$ (population of the upper level), $A_{ki}$ it the transition probability for transitions from upper level $k$ to lower level $i$, and $h\nu_{ik}$ is the photon energy (or the energy difference between the upper level and lower level). Although both $A_{ki}$ and $\nu_{ik}$ are well defined quantities for each line of a given atom, the population values $N_k$ depend on plasma conditions in a given light source, and they are this different for different sources.

Taking into account this issue, the following points should be kept in mind when using relative intensities:

1. There is no common scale for relative intensities. The values from different databases or different publications use different scales. The relative intensities have meaning only within a given spectrum.
2. The relative intensities are most useful in comparing strengths of spectral lines that are not separated widely. This results from the fact that most relative intensities are not corrected for spectral sensitivity of the measuring instruments.

3. Relative intensities are source dependent (either laboratory or astrophysical detections)
4. When an instance of the model includes multiple `Line.intensity` values, all of them must have the same scale.

In general, the `intensity` will represent an integrated flux fitted to a specific model for the line.

### 3.3.14 Line.observedFlux: PhysicalQuantity
Integrated flux of the line profile over a given wavelength range

### 3.3.15 Line.observedFluxWaveMin: PhysicalQuantity
Minimum wavelength for observedFlux integration.

### 3.3.16 Line.observedFluxWaveMax: PhysicalQuantity

Maximum wavelength for observedFlux integration.

### 3.3.17 Line.significanceOfDetection: PhysicalQuantity

The significance of line detection in an observed spectrum. It can be expressed in terms of signal-to-noise ratio, or detection probability (usually null hypothesis probability that a given observed line is due to a statistical background fluctuation). Usually represents the difference between best-fit flux and zero flux.

### 3.3.18 Line.transitionType: String

String indicating the first non zero term in the expansion of the operator $e^{i\vec{k}\vec{r}}$ in the atomic transition probability integral:

$$\int \phi^f e^{i\vec{k}\vec{r}} l \nabla \phi_i d^3 x$$

Possible values correspond to, e.g., "electric dipole", "magnetic dipole", "electric quadrupole", etc., or their corresponding common abbreviations E1, M1, E2, etc.

### 3.3.19 Line.strength: PhysicalQuantity

In theoretical works, the line strength S is widely used (Drake 1996):

$$S = S(i,k) = S(k,i) = |R_{ik}|^2 \quad where \quad R_{ik} = \langle \psi_k | P | \psi_i \rangle$$

Where $\psi_i$ and $\psi_k$ are the initial- and final-state wavefunction and $R_{ik}$ is the transition matrix element of the appropriate multipole operator $P$. For example, the relationship between $A$, $f$, and $S$ for electric dipole ($E1$ or allowed) transitions in S.I. units ($A$ in s$^{-1}$, $\nu$ in s$^{-1}$, $S$ in m$^2$ C$^2$, $\varepsilon_0$ in C$^2$.N$^{-1}$.m$^{-2}$, $h$ in J.s) are:

$$A_{ik} = \left( \frac{4\pi h \alpha \nu^2}{c^2 m_e} \right) \left( \frac{g_k}{g_i} \right) f_{ki} = \left( \frac{16\pi^3 \nu^3}{3h\varepsilon_0 c^3 g_i} \right) S$$

### 3.3.20 Line.observedBroadeningCoefficient: PhysicalQuantity

Line broadening can be divided into *natural line broadening*, which is always present, *Doppler line broadening*, due to the motion of the observed atoms in in different directions with different velocities with the resulting Doppler shifts producing a spread in the frequency of the observed line, and *collisional* or *pressure line broadening*, due to the effects of other particles on the radiating atom.

The `observedBroadeningCoefficient` represents the width of the line profile at half maximum (FWHM, Full Width Half Maximum induced by a process of `Process.type="Broadening"` (see `Process` class below).

### 3.3.21 Line.observedShiftingCoefficient: PhysicalQuantity

Shift of the transition laboratory wavelength induced by a process of `Process.type="Energy shift"` (see `Process` class below). It is expressed by the difference between the peak intensity wavelength in the observed profile and the laboratory wavelength.

## 3.4 Species

This class is a placeholder for a future model, providing a full description of the physical and chemical property of the chemical element of compound where the transition originating the line occurs.
While a more detailed data model for this class is built, this simplified version contains simply the name of the Species involved in the transition.

### 3.4.1 Species.name: String

Name of the chemical element or compound including ionisation status.

Examples:

"`CIV`" for Carbon three times ionised
"`N2H+`" for the Dyazenylium molecule

(see Appendix A for standard chemical element names).

## 3.5  Level

When an atom or a molecule is unperturbed by external fields, its Hamiltonian commutes with the resultant angular momentum **J**. As a consequence, an energy level corresponding to a given value of *j* is (2*j*+1) degenerate, each of the different states being characterised by a different value of the magnetic moment *m*. Such a set of (2j+1) states we call a `Level`. When an atom or molecule is perturbed and the different states do not all have the same energy it is still convenient to refer to such a set of states by a generic name, so we use `Level` in this wider sense. This of this class is to describe the quantum mechanical properties of each level, between which the transition originating the line occurs.

### 3.5.1  Level.totalStatWeight: PhysicalQuantity

An integer representing the total number of terms pertaining to a given level. For instance, in an LS coupling scheme (see Appendix B) the total statistical weight of all levels (L,S)J belonging to a given term LS is: *g*=(2*L*+1)(2*S*+1). For a single electron with specified *n*, the statistical weight is $2n^2$.

### 3.5.2  Level.nuclearStatWeight: PhysicalQuantity

The same as `Level.totalStatWeight` for nuclear spin states only

### 3.5.3  Level.landeFactor: PhysicalQuantity

A dimensionless factor *g* that accounts for the splitting of normal energy levels into uniformly spaced sublevels in the presence of a magnetic field. The level of energy $E_0$ is split into levels of energy:

$$E_0 + g\beta B(J), E_0 + g\beta B(J-1), \ldots, E_0 - g\beta B(J)$$

Where *B* is the magnetic field and *β* is a proportionality constant.

In the case of the L-S coupling (see Appendix B), the Lande factor $g_j$ is specified as the combination of atomic quantum numbers, which enters in the definition of the total magnetic moment $m_j$ in the fine structure interaction:

$$m_j = \frac{g_j \mu_B J}{\hbar}$$

where $\mu_B$ is the Bohr magneton, defined as:

$$\mu_B = \frac{eh}{2\pi m_e}$$

Where

$e$ is the elementary charge

$h$ is the Planck constant

$m_e$ is the electron rest mass

In terms of pure quantum numbers:

$$g_j(J,L,S) \equiv 1 + \frac{J(J+1) + S(S+1) + L(L+1)}{2J(J+1)}$$

### 3.5.4  Level.lifeTime: PhysicalQuantity
Intrinsic lifetime of a level due to its radiative decay.

### 3.5.5  Level.energy: PhysicalQuantity
The binding energy of an electron belonging to the level.

### 3.5.6  Level.energyOrigin: String
Human readable string indicating the nature of the energy origin. Examples: "Ionization energy limit", "Ground state energy" of an atom, "Dissociation limit" for a molecule, etc

### 3.5.7  Level.quantumState: QuantumState
A representation of the level quantum state through its set of quantum numbers

### 3.5.8 Level.nuclearSpinSymmetryType: String

A string indicating the type of nuclear spin symmetry. Possible values are: `"para"`, `"ortho"`, `"meta"`

### 3.5.9 Level.parity: PhysicalQuantity

Eigenvalue of the parity operator. Values (+1,-1)

### 3.5.10 Level.configuration: String

For atomic levels, the standard specification of the quantum numbers *nPrincipal* (*n*) and **lElectronicOrbitalAngularMomentum** (*l*) for the orbital of each electron in the level; an exponent is used to indicate the numbers of electrons sharing a given *n* and *l*. For example, $1s^2, 2s^2, 2p^6, 5f$. The orbitals are conventionally listed according to increasing *n*, then by increasing *l*, that is, 1*s*, 2*s*, 2*p*, 3*s*, 3*p*, 3*d*, ….. Closed shell configurations may be omitted from the enumeration (see section 4 for Quantum Number definitions).

For molecular states, similar enumerations takes place involving appropriate representations.

## 3.6 QuantumState

In Quantum Mechanics, a pure state is represented by a bundle $|\Psi(t)\rangle_B$ of a complex and separable Hilbert space to which the physical system is associated. An element $|\Psi(t)\rangle$ of that bundle is called the *state vector*.

The state vectors of a system evolve in time following the Schrödinger equation:

$$i\hbar \frac{\partial}{\partial t}|\Psi(t)\rangle = H(t)|\Psi(t)\rangle$$

where *H(t)* is the *Hamiltonian* of the system (also called the *energy operator* in cases where $H \neq H(t)$ ). The system's observables are represented by operators which are constant in time. In the stationary states, where the Hamiltonian is not a function of time, then $|\Psi(t_0)\rangle$ is an *eigenvector* of the Hamiltonian *H* of *eigenvalue E* such that:

$$H|\Psi(t_0)\rangle = E|\Psi(t_0)\rangle$$

For the particular case of atoms for instance, the state of the atom can be specified by using 4 quantum numbers (4N quantum numbers for N electrons):

$n = 1, 2, .....$ = principal quantum number  
$l = 0, 1, ...., n-1$ = orbital angular momentum  
$m_l = -l, ...., +l$ = magnetic quantum number  
$m_s = -1/2, +1/2$ = spin quantum number

(see below for further details on Quantum Numbers)

### 3.6.1 QuantumState.mixingCoefficient: PhysicalQuantity

A positive or negative number giving the squared or the signed linear coefficient corresponding to the associated component in the expansion of the eigenstate (QuantumState in the DM). It varies from 0 to 1 (or -1 to 1)

### 3.6.2 QuantumState.quantumNumber: QuantumNumber

In order to allow for a simple mechanism for quantum numbers coupling, the QuantumNumber object is reduced to the minimum set of needed attributes to identify a quantum number. Coupling is then implemented by specifying combinations of the different quantum numbers.

### 3.6.3 QuantumState.termSymbol: String

The term (symbol) to which this quantum state belongs, if applicable.

For example, in the case of Spin-Orbit atomic interaction, a term describes a set of $(2S+1)(2L+1)$ states belonging to a definite configuration and to a definite L and S. The notation for a term is for the LS coupling is, at follows:

$$^{2S+1}L_J$$

where

- $S$ is the total spin quantum number. $2S+1$ is the **spin multiplicity**: the maximum number of different possible states of $J$ for a given ($L$,$S$) combination.
- $L$ is the total orbital quantum number in spectroscopic notation. The symbols for L = 0,1,2,3,4,5 are S,P,D,F,G,H respectively.
- $J$ is the total angular momentum quantum number, where $|L-S| \leq J \leq L+S$

For instance, $^3P_1$ would describe a term in which L=1, S=1 and J=1. If J is not present, this term symbol represents the 3 different possible levels (J=0,1,2)

See Appendix B for more examples of different couplings.

For molecular quantum states, it is a shorthand expression of the group irreducible representation and angular momenta that characterize the state of a molecule, i.e its electronic quantum state. A complete description of the molecularTermSymbol can be found in « Notations and Conventions in Molecular Spectroscopy: Part 2. Symmetry notation » (IUPAC Recommendations 1997), C.J.H. Schutte et at, Pure & Appl. Chem., Vol. 69, no. 8, pp. 1633-1639, 1997. The molecular term symbol contains the irreducible representation for the molecular point groups with right subscripts and superscripts, and a left superscript indicating the electron spin multiplicity, Additionally it starts with an symbol ~X (i.e., ~ on X) (ground state), Ã, ~B (i.e. ~ on B), ... indicating excited states of the same multiplicity than the ground state X or ã, ~b (~ on b), ... for excited states of different multiplicity.

## 3.7 QuantumNumber

The scope of this class is to represent the set of quantum numbers describing each `QuantumState`.

### 3.7.1 QuantumNumber.label: String

The name of the quantum number. It is a string like "F", "J", "I1", etc., or whatever human readable string that identifies the quantum number

### 3.7.2 QuantumNumber.type: String

A string describing the quantum number. Recommended values are (see Chapter 4 for a description):

```
totalNuclearSpinI
totalMagneticQuantumNumberI
totalMolecularProjectionI
nuclearSpin
parity
serialQuantumNumber
nPrincipal
lElectronicOrbitalAngularMomentum
sAngularMomentum
jTotalAngularMomentum
fTotalAngularMomentum
lMagneticQuantumNumber
sMagneticQuantumNumber
jMagneticQuantumNumber
fMagneticQuantumNumber
asymmetricTAU
asymmetricKA
asymmetricKC
totalSpinMomentumS
totalMagneticQuantumNumberS
totalMolecularProjectionS
totalElectronicOrbitalMomentumL
totalMagneticQuantumNumberL
totalMolecularProjectionL
totalAngularMomentumN
totalMagneticQuantumNumberN
totalMolecularProjectionN
totalAngularMomentumJ
totalMagneticQuantumNumberJ
totalMolecularProjectionJ
intermediateAngularMomentumF
totalAngularMomentumF
totalMagneticQuantumNumberF
vibrationNu
vibrationLNu
totalVibrationL
vibronicAngularMomentumK
vibronicAngularMomentumP
hinderedK1
hinderedK2
```

### 3.7.3 QuantumNumber.numeratorValue: PhysicalQuantity

The numerator of the quantum number value

### 3.7.4  QuantumNumber.denominatorValue: PhysicalQuantity

The denominator of the quantum number value. If not explicitly specified, it is defaulted to "1" (meaning that the corresponding quantum number value is a multiple integer)

### 3.7.5  QuantumNumber.description: String

A human readable string, describing the nature of the quantum number. Standard descriptions are given in Chapter 4 for those quantum numbers whose names are given above.  For a quantum number not appearing above, the description shall be given here.

## 3.8 Process

The scope of this class is to describe the physical process responsible for the generation of the line, or for the modification of its physical properties with respect to those measured in the laboratory. The complete description of the process is relegated to specific placeholder called "model" which will describe specific physical models for each process.

### 3.8.1 Process.type: String

String identifying the type of process. Possible values are: "Matter-radiation interaction", "Matter-matter interaction", "Energy shift", "Broadening".

### 3.8.2 Process.name: String

String describing the process: Example values (corresponding to the values of "type" listed above) are: "Photoionization", "Collisional excitation", "Gravitational redshift", "Natural broadening".

### 3.8.3 Process.model: PhysicalModel

A theoretical model by which a specific process might be described.

## 3.9 Environment

The scope of this class is describing the physical properties of the ambient gas, plasma, dust or stellar atmosphere where the line is generated.

### 3.9.1 Environment.temperature: PhysicalQuantity

The temperature in the line-producing plasma.

### 3.9.2 Environment.opticalDepth: PhysicalQuantity

The optical depth, $\tau_v$, is defined as:

$$d\tau_v = \kappa_v \rho dx$$

where ρ is the density of plasma, and the absorption coefficient $\kappa_v$ is defined as the fractional decrease in intensity per unit distance:

$$\frac{dI}{I} = -\kappa_\rho dx$$

### 3.9.3 Environment.particleDensity: PhysicalQuantity

The particle density in the line-producing plasma.

### 3.9.4 Environment.massdensity: PhysicalQuantity

The mass density in the line-producing plasma.

### 3.9.5 Environment.pressure: PhysicalQuantity

The pressure in the line-producing plasma.

### 3.9.6 Environment.entropy: PhysicalQuantity

The entropy of the line-producing plasma.

### 3.9.7 Environment.mass: PhysicalQuantity

The total mass of the line-producing gas/dust cloud or star.

### 3.9.8 Environment.metallicity: PhysicalQuantity

As customary in astronomy, the metallicity of an element is expressed as the logarithmic ratio between the element and the Hydrogen abundance, normalized to the solar value. If the metallicity of a celestial object or plasma is expressed through a single number, this refers to the iron abundance.

### 3.9.9 Environment.extinctionCoefficient: PhysicalQuantity

A quantitative observable $k$, which expresses the suppression of the emission line intensity due to the presence of optically thick matter along the line-of-sight. It is a measure of the intervening gas density through one of the following equations:

$$k = n\sigma = \kappa\rho$$

where $n$ is the particle density, $\sigma$ is the integrated cross section, $\kappa$ is the integrated opacity and $\rho$ the matter density.

### 3.9.10 Environment.model: PhysicalModel

Placeholder for future detailed theoretical models of the environment plasma where the line appears.

## 3.10 Source

This class gives a basic characterization of the celestial source, where an astronomical line has been observed. The Source Data Model should be developed further within the IVOA context. For the time being, minimal information about the Source is given.

### 3.10.1 Source.IAUname: String

The IAUname of the source

### 3.10.2 Source.name: String

An alternative or conventional name of the source

### 3.10.3 Source.coordinates: STC

Coordinates of the source. Link to IVOA Space Time Coordinates data model

# 4  Quantum Numbers

The list contains the most usual quantum numbers in atomic and molecular spectroscopy. The list is not exhaustive and is opened to new entries.

Note for molecules: Angular momemtum basis functions, $|A\ \alpha\ M_A>$, can be simultaneous eigenfunctions of three types of operators : the magnitude $A^2$, the component of $A$ onto the internuclear axis $A_z$, and the component of $A$ on the laboratory quantization axis $A_Z$. The basis function labels A, $\alpha$ and $M_A$ correspond to eigenvalues of $A^2$, $A_z$, and $A_Z$, respectively $\hbar^2 A(A+1)$, $\hbar\ \alpha$ and $\hbar\ M_A$.

Note for intermediate coupling: Intermediate coupling occurs in both atomic and molecular physics. The document below gives some explanations about intermediate coupling in atomic physics, these explanations can be transposed to molecular physics (as for intermediate coupling between different Hund's cases). As described below, levels can be labelled by the least objectionable coupling case, by linear combinaison of pure coupling basis functions (the linear coefficients can be determined in a theoretical approach: *this is planned for in the data model*), or simply by a sort of serial number (see **serialQuantumNumber**  below)

## 4.1 Various Quantum numbers

### 4.1.1 totalNuclearSpinI
total nuclear spin of one atom or a molecule, I

### 4.1.2 totalMagneticQuantumNumberI
total magnetic quantum number, $M_I = -I, -I+1, ..., I-1, I$ where $\hbar M_I$ is the eigenvalue of the $\hat{I}_z$ operator

### 4.1.3 totalMolecularProjectionI
total nuclear spin projection quantum number $\Omega_I$, $\Omega_I = -I, -I+1, ..., I-1, I$ where $\hbar \Omega_I$ is the eigenvalue of the $\hat{I}_z$ operator

### 4.1.4 nuclearSpin
nuclear spin of individual nucleus $i$ which composes a molecule, noted; $I_i$

### 4.1.5 parity
eigenvalue of the parity operator applied to the total wavefunction. It takes the value "0" for even parity and "1" for odd parity

### 4.1.6 serialQuantumNumber
A serial quantum number that labels states to which no good or nearly good quantum numbers can be assigned to.

## 4.2 Quantum numbers for hydrogenoids

### 4.2.1 nPrincipal
principal quantum number n

### 4.2.2 lElectronicOrbitalAngularMomentum
orbital angular momentum of an electron $l = 0,1,2,...$ where $\hbar^2(l+1)$ is the eigenvalue of the $\hat{l}^2$ operator (called as well azimuthal quantum number).

### 4.2.3 sAngularMomentum
spin angular momentum of an electron, $s = \frac{1}{2}$ only where $\hbar^2 s(s+1)$ is the eigenvalue of the $\hat{s}^2$ operator)

### 4.2.4 jTotalAngularMomentum
total angular momentum of one electron $j$, $j = l - 1/2 (l > 0)$ and $j = l + 1/2$. $\hbar^2 j(j+1)$ is the eigenvalue of the $\hat{j}^2$ operator, where $\hat{j} = \hat{l} + \hat{s}$

### 4.2.5 fTotalAngularMomentum
total angular momentum $f$, including nuclear spin $I$. $\hbar^2 f(f+1)$ is the eigenvalue of the $\hat{f}^2$ operator, where $\hat{f} = \hat{j} + \hat{I}$

### 4.2.6 lMagneticQuantumNumber
orbital magnetic quantum number, $m_l = -l, -l+1, ..., l-1, l$ where $\hbar m_l$ is the eigenvalue of the $\hat{l}_z$ operator

### 4.2.7 sMagneticQuantumNumber
spin magnetic quantum number, $m_s = \pm 1/2$ where $\hbar m_s$ is the eigenvalue of the $\hat{s}_z$ operator

### 4.2.8 jMagneticQuantumNumber

orbital magnetic quantum number, $m_j = -j, -j+1, ..., j-1, j$ where $\hbar m_j$ is the eigenvalue of the $\hat{j}_z$ operator

### 4.2.9 fMagneticQuantumNumber

orbital magnetic quantum number, $m_f = -f, -f+1, ..., f-1, f$ where $\hbar m_f$ is the eigenvalue of the $\hat{f}_z$ operator

## 4.3 Pure rotational quantum numbers

### 4.3.1 asymmetricTAU
Index $\tau$ labelling asymmetric rotational energy levels for a given rotational quantum number N.

Note: The solution of the Schrödinger equation for an asymmetric-top molecule gives for each value of N, (2N+1) eigenfunctions with its own energy. It is customary to keep track of them by adding the subscript $\tau$ to the N value ($N_\tau$). This index $\tau$ goes from -N for the lowest energy of the set to +N for the highest energy, and is equal to $(K_a - K_c)$.

### 4.3.2 asymmetricKA
For a given N, energy levels may be specified by $K_a$ $K_c$ (or $K_{-1}$ $K_1$, or $K_-$ $K_+$ are alternative notations), where $K_a$ is the K quantum number for the limiting prolate (B=C) and $K_c$ for the limiting oblate (B=A). In the notation ($K_{-1}$ $K_1$) the subscripts "1" and "-1" correspond to values of the asymmetry parameter $\kappa = \dfrac{2B - A - C}{A - C}$ where A, B, C are rotational constants of the asymmetric molecule (by definition A>B>C)

### 4.3.3 asymmetricKc
see **asymmetricKA**

## 4.4 Quantum numbers for n electron systems (atoms and molecules)

### 4.4.1 totalSpinMomentumS
it is the total spin quantum number S, S can be integral or half-integral. $\hbar^2 S(S+1)$ is the eigenvalue of the $\hat{S}^2$ operator, where $\hat{S} = \Sigma_{i=1}^{N}\hat{s}_i$

### 4.4.2 totalMagneticQuantumNumberS
total spin magnetic quantum number, $M_S = -S, -S+1, ..., S-1, S$ where $\hbar M_S$ is the eigenvalue of the $\hat{S}^2$ operator.

### 4.4.3 totalMolecularProjectionS
total spin projection quantum number $\Sigma$, $\Sigma = -S, -S+1, ..., S-1, S$, where $\hbar\Sigma$ is the eigenvalue of the $\hat{S}^2$ operator.

### 4.4.4 totalElectronicOrbitalMomentumL
it is the total orbital angular momentum $L$, $L$ is integral. $\hbar^2 L(L+1)$ is the eigenvalue of the $\hat{L}^2$ operator, where $\hat{L} = \Sigma_{i=1}^{n}\hat{l}_i$

### 4.4.5 totalMagneticQuantumNumberL
total orbital magnetic quantum number, $M_L = -L, -L+1, ..., L-1, L$, where $\hbar M_L$ is the eigenvalue of the $\hat{L}_z$ operator.

### 4.4.6 totalMolecularProjectionL
total orbital projection quantum number, $\Lambda = 0, 1, ..., L-1, L$ where $\hbar\Lambda$ is the absolute value of the eigenvalue of the $\hat{L}_z$ operator (Hund's cases (a) and (b) in the case of a diatomic)

### 4.4.7 totalAngularMomentumN

is the total angular momentum N exclusive of nuclear and electronic spin, N is integral. For a molecule in a close-shell state **totalAngularMomemtumN** is the pure rotational angular momentum.

### 4.4.8 totalMagneticQuantumNumberN

total orbital magnetic quantum number, $M_N = -N, -N+1, ..., N-1, N$ where $\hbar M_N$ is the eigenvalue of the $\hat{N}_Z$ operator

### 4.4.9 totalMolecularProjectionN

absolute value of the component of the angular momentum $N$ along the axis of a symmetric (or quasi-symmetric) rotor, usually noted $K$. $\pm \hbar K$ is the eigenvalue of the $\hat{N}_Z$ operator, with values $K = 0, ..., N-1, N$

For open shell diatomic molecules, it corresponds to "totalMolecularProjectionL" ($\Lambda$), so we advise to preferentially use "totalMolecularProjectionL"

Note: The symbol $K$ is also used in spectroscopy to describe the component of the vibronic angular momentum (excluding spin) along the axis for linear polyatomic molecules. In this model, we prefer to uniquely identify this specific case by a different type of Quantum Number: "vibronicAngularMomentumK", defined thereafter.

### 4.4.10 totalAngularMomentumJ

is the total angular momentum J exclusive of nuclear spin, J can be integral or half-integral.

For atoms:
$$\hat{J} = \Sigma_{i=1}^{n}(\hat{l}_i + \hat{s}_i)$$

For molecules:

$$\hat{J} = \hat{N} + \Sigma_{i=1}^{n}\hat{s}_i = \hat{N} + \hat{S}$$

### 4.4.11 totalMagneticQuantumNumberJ

total magnetic quantum number, $M_J = -J, -J+1, ..., J-1, J$ where $\hbar M_J$ is the eigenvalue of the $\hat{J}_Z$ operator.

### 4.4.12 totalMolecularProjectionJ

absolute value of the component of the angular momentum $J$ along the molecular axis, noted $\Omega = 0, ..., J-1, J$ where $\pm\hbar\Omega$ is the eigenvalue of the $\hat{J}_z$ operator.

For linear molecules with no nuclear spin (or no nuclear spin coupled to the molecular axis), it is the absolute value of the component of the total electronic angular momentum $\hat{L} + \hat{S}$ on the molecular axis (Hund's cases (a) and (c)). When $\Lambda$ and $\Sigma$ are defined (Hund's case (a)): $\Omega = |\Lambda \pm \Sigma|$

For linear molecules with a nuclear spin coupled to the molecular axis, it includes as well the component $\Omega_I$ of the nuclear spin on the molecular axis.

### 4.4.13 intermediateAngularMomemtumF

is associated to the intermediate quantum number $F_i$ where $\hat{F}_i = \hat{I}$ (or $\hat{I}_j$) + any other vector

### 4.4.14 totalAngularMomentumF

is the total angular momentum $F$ including nuclear spin, $F$ can be integral or half-integral. $F(F+1)\hbar^2$ is the eigenvalue of the $\hat{F}^2$ operator, where for atoms:

$$\hat{F} = \Sigma_{i=1}^{n}(\hat{l}_i + \hat{s}_i) + \hat{I} = \hat{J} + \hat{I}$$

and for molecules with *m* nuclear spins:

$$\hat{F} = \hat{J} + \Sigma_{i=1}^{n}\hat{I}_i = \hat{J} + \hat{I}$$

### 4.4.15 totalMagneticQuantumNumberF

total magnetic quantum number, $M_F = -F, -F+1, ..., F-1, F$ where $\hbar M_F$ is the eigenvalue of the $\hat{F}_Z$ operator

## 4.5 Vibrational and (ro-)vibronic quantum numbers

### 4.5.1 vibrationNu

vibrational modes $v_i$ (following Mulliken conventions). By default the vibrational mode is a normal mode. If the vibrational mode is fairly localised, the bond description will be included in the attribute "description" of "QuantumNumber"

### 4.5.2 vibrationLNu

angular momentum associated to degenerate vibrations, $l_i = v_i, v_i - 2, v_i - 4, ..., 1$ or $0$

### 4.5.3 totalVibrationL

total vibrational angular momentum is the sum of all angular momenta $\pm l_i$ associated to degenerate vibrations: $l_v = |\Sigma_i(\pm l_i)|$

### 4.5.4 vibronicAngularMomentumK

is the sum of the total vibrational angular momentum $l_v$ and of the electronic orbital momentum about the internuclear axis $\Lambda$: $K = |\pm l_v \pm \Omega|$ here $K$ ($l_v$ and $\Lambda$) are unsigned quantities. This is used for linear polyatomic molecules. (see p.25 Volume III pf Herzberg, and REC. (recommendation) 17 of Muliken, 1955)

### 4.5.5 vibronicAngularMomentumP

is the sum of the total vibrational angular momentum $l_v$ and of the total electronic orbital momentum about the internuclear axis $\Omega$: $P = |\pm l_v \pm \Omega|$ here $P$ ($l_v$ and $\Omega$) are unsigned quantities. This is used for linear polyatomic molecules. (see p.26 Volume III pf Herzberg, and REC. 18 of Muliken, 1955)

### 4.5.6 rovibronicAngularMomentumP

total resultant axial angular momentum quantum number including electron spin:

$P = |K + \Sigma|$. (REC. 26 of Mulliken, 1955)

### 4.5.7 hinderedK1, hinderedK2

for internal free rotation of 2 parts of a molecule, (see p.492, Volume II of Herzberg), 2 additional projection quantum numbers are necessary: $k_1$ and $k_2$, such that total rotational energy is given by:

$$F(N, K, k_1, k_2) = BN(N+1)BK^2 + A_1 k_1^2 + A_2 k_2^2$$

Where $N$ is **totalAngularMomentumN** and $K$ is **totalMolecularProjectionN** (see p.492, Volume II of Herzberg (1964))

# 5 Attributes table

| Attribute | Type | Meaning |
|---|---|---|
| PhysicalQuantity | Class | Class representing a Physical Quantity |
| PhysicalQuantity.value | Real Double Integer | Value of a Physical Quantity |
| PhysicalQuantity.error | Real Double Integer | Error associated to a PhysicalQuantity value. Both value and error must be in same units |
| PhysicalQuantity.unit | Unit | Unit class associated to the PhysicalQuantity |
| Unit | Class | Class representing a Unit |
| Unit.expression | String | Human readable string representation of the unit |
| Unit.scaleSI | Double | Conversion scale factor to International System of units |
| Unit.dimEquation | String | Representation of the unit dimensions |
| Line | Class | Class representing a Spectroscopic line |
| Line.title | String | Human readable description of the line |
| Line.initialLevel | Level | Initial level of the transition originating the line |
| Line.finallevel | Level | Final level of the transition originating the line |
| Line.initialElement | Species | Description of the initial state of the atom or molecule |
| Line.finalElement | Species | Description of the final state of the atom or molecule |
| Line.wavelength | PhysicalQuantity | Wavelength in vacuum of the transition originating the line |
| Line.frequency | PhysicalQuantity | Frequency in vacuum of the transition originating the line |

| Attribute | Type | Meaning |
| --- | --- | --- |
| Line.wavenumber | PhysicalQuantity | Wavenumber in vacuum of the transition originating the line |
| Line.airWavelength | PhysicalQuantity | Wavelength in the air of the transition originating the line |
| Line.einsteinA | PhysicalQuantity | Einstein coefficient for spontaneous emission from Final to Initial levels |
| Line.oscillatorStrength | PhysicalQuantity | Historical representation of the strength of a transition related to pre quantum-mechanics interpretation of electromagnetic waves as harmonic oscillators (related to Einstein A coeff.) |
| Line.weightedOscillatorStrength | PhysicalQuantity | Product between oscillator strength and statistical weight of the initial level |
| Line.intensity | PhysicalQuantity | Integrated flux over a fitted model to the line |
| Line.observedFlux | PhysicalQuantity | Integrated flux over a given wavelength range |
| Line.observedFluxWaveMin | PhysicalQuantity | Minimum wavelength for observedFlux integration |
| Line.observedFluxWaveMax | PhysicalQuantity | Maximum wavelength for observedFlux integration |
| Line.significanceOfDetection | PhysicalQuantity | The significance of line detection in an observed spectrum |
| Line.transitionType | String | Type of the transition, e.g., e.g., "electric dipole", "magnetic dipole", "electric quadrupole", etc., or "E1", "M1", "E2", etc. |
| Line.strength | PhysicalQuantity | Quantum mechanical quantity representing line strength (related to Einstein A coeff. and oscillatorStrength) |

| Attribute | Type | Meaning |
| --- | --- | --- |
| Line.observedBroadeningCoefficient | PhysicalQuantity | Observed broadening of a line due to Natural, Doppler, or collisional broadening processes |
| Line.observedShiftingCoefficient | PhysicalQuantity | Shift of the transition laboratory wavelength induced by a process of type Energy Shift |
| Species | Class | Placeholder Class for a future model, providing a full description of the physical and chemical property of the chemical element of compound where the transition originating the line occurs. |
| Species.name | String | Name of the chemical element or compound including ionisation status |
| Level | Class | Class representing Initial and Final levels between which the trnasition happens |
| Level.totalStatWeight | PhysicalQuantity | An integer representing the total number of terms pertaining to a given level |
| Level.nuclearStatWeight | PhysicalQuantity | The same as Level.totalStatWeight for nuclear spin states only |
| Level.landeFactor | PhysicalQuantity | A dimensionless factor g that accounts for the splitting of normal energy levels into uniformly spaced sublevels in the presence of a magnetic field |
| Level.lifeTime | PhysicalQuantity | Intrinsic lifetime of a level due to its radiative decay |
| Level.energy | PhysicalQuantity | The binding energy of an electron belonging to the level |

| Attribute | Type | Meaning |
| --- | --- | --- |
| Level.energyOrigin | String | Human readable string indicating the nature of the energy origin, e.g., "Ionization energy limit", "Ground state energy" of an atom, "Dissociation limit" for a molecule, etc |
| Level.quantumState | QuantumState | A representation of the level quantum state through its set of quantum numbers |
| Level.nuclearSpinSymmetryType | String | A string indicating the type of nuclear spin symmetry. Possible values are: "para", "ortho", "meta" |
| Level.parity | PhysicalQuantity | Eigenvalue of the parity operator. Values (+1,-1) |
| Level.configuration | String | Human readable string representing the corresponding level configuration |
| QuantumState | Class | Class representing the Quantum State of the corresponding level in the transition |
| QuantumState.mixingCoefficient | PhysicalQuantity | A positive or negative number giving the squared or the signed linear coefficient corresponding to the associated component in the expansion of the eigenstate |
| QuantumState.quantumNumber | QuantumNumber | Quantum number(s) describing the state |
| QuantumState.termSymbol | String | The term (symbol) to which this quantum state belongs, if applicable |
| QuantumNumber | Class | Class representing the quantum numbers describing each quantum state |

| Attribute | Type | Meaning |
| --- | --- | --- |
| QuantumNumber.label | String | Human readable string representing the name of the quantum number. It is a string like "F", "J", "I1", etc., |
| QuantumNumber.type | String | String representing the quantum number, recommended within this model (c.f. Chapter 4) |
| QuantumNumber.numeratorValue | PhysicalQuantity | The numerator of the quantum number value. For non half-integers, this number would just represent the value of the quantum number |
| QuantumNumber.denominatorValue | PhysicalQuantity | The denominator of the quantum number value. It defaults to 1 (one) if not explicitly input (as would be the case for non half-integer numbers) |
| QuantumNumber.description | String | A human readable string, describing the nature of the quantum number |
| Process | Class | Class representing the physical process responsible for the generation of the line |
| Process.type | String | String identifying the type of process. Possible values are: "Matter-radiation interaction", "Matter-matter interaction", "Energy shift", "Broadening" |
| Process.name | String | String describing the process: Example values (corresponding to the values of "type" listed above) are: "Photoionization", "Collisional excitation", "Gravitational redshift", "Natural broadening". |
| Process.model | PhysicalModel | A theoretical model by which a specific process might be described |

| Attribute | Type | Meaning |
| --- | --- | --- |
| Environment | Class | Class representing the physical properties of the ambient gas, plasma, dust or stellar atmosphere where the line is generated |
| Environment.temperature | PhysicalQuantity | The temperature in the line-producing plasma |
| Environment.opticalDepth | PhysicalQuantity | The optical depth of the line producing plasma |
| Environment.particleDensity | PhysicalQuantity | The particle density in the line producing plasma |
| Environment.massdensity | PhysicalQuantity | The mass density in the line-producing plasma |
| Environment.pressure | PhysicalQuantity | The pressure in the line producing plasma |
| Environment.entropy | PhysicalQuantity | The entropy of the line producing plasma |
| Environment.mass | PhysicalQuantity | The total mass of the line-producing gas/dust cloud or star |
| Environment.metallicity | PhysicalQuantity | The logarithmic ratio between the element and the Hydrogen abundance, normalized to the solar value |
| Environment.extinctionCoefficient | PhysicalQuantity | The suppression of the emission line intensity due to the presence of optically thick matter along the line-of-sight |
| Environment.model | PhysicalModel | Placeholder for future detailed theoretical models of the environment plasma where the line appears. |
| Source | Class | Class representing a basic characterization of the celestial source, where an astronomical line has been observed |
| Source.IAUname | String | The IAUname of the source |

| Attribute | Type | Meaning |
|---|---|---|
| Source.name | String | An alternative or conventional name of the source |
| Source.coordinates | STC | Coordinates of the source. Link to IVOA Space Time Coordinates data model |

# 6 UCDs

The following is a list of the UCDs that should accompany any of the object attributes in their different serializations.

They are based in *"[The UCD1+ controlled vocabulary Version 1.23](#)"* (IVOA Recommendation, 2 Apr 2007).

There is one table per each of the objects in the Data Model. The left column indicates the object attribute, and the right column the UCD. Items appearing in *(bold)* correspond to other objects in the model.

| Line | |
|---|---|
| initialLevel | *(Level)* |
| finalLevel | *(Level)* |
| initialElement | *(ChemicalElement)* |
| finalElement | *(ChemicalElement)* |
| wavelength | em.wl |
| wavenumber | em.wn |
| frequency | em.freq |
| airWavelength | em.wl |
| einsteinA | phys.at.transProb |
| oscillatorStrength | phys.at.oscStrength |
| weightedOscillStrength | phys.at.WOscStrength |
| intensity | spect.line.intensity |
| observedFlux | phot.flux |
| observedFluxWaveMin | em.wl |
| observedFluxWaveMax | em.wl |
| significanceOfDetection | stat.snr |
| process | *(Process)* |
| lineTitle | meta.title |
| transitionType | meta.title |
| strength | spect.line.strength |
| observedBroadeningCoefficient | spect.line.broad |
| observedShiftingCoefficient | phys.atmol.lineShift |

| Species | |
|---|---|
| name | meta.title |

| Level | |
|---|---|
| type | meta.title |
| totalStatWeight | phys.atmol.sweight |
| nuclearStatWeight | phys.atmol.nucweigth |
| lifeTime | phys.atmol.lifetime |
| energy | phys.energy |
| quantumState | *(QuantumState)* |
| energyOrigin | phys.energy |
| landeFactor | phys.at.lande |
| nuclearSpinSymmetryType | phys.atmol.symmetrytype |
| parity | phys.atmol.parity |
| energyOrigin | phys.energy |
| configuration | phys.atmol.configuration |

| QuantumState | |
|---|---|
| normalizedProbability | stat.normalProb |
| quantumNumber | phys.atmol.qn |
| termSymbol | phys.atmol.termSymbol |

| QuantumNumber | |
|---|---|
| label | meta.title |
| type | meta.title |
| numeratorValue | meta.number |
| denominatorValue | meta.number |
| description | meta.note |
|  |  |

| Process | |
|---|---|

| model | *(Model)* |
|---|---|
| name | meta.title |

| Environment ||
|---|---|
| temperature | phys.temperature |
| opticalDepth | phys.absorption.opticalDepth |
| density | phys.density |
| pressure | phys.pressure |
| extinctionCoefficient | phys.absorption |
| entropy | phys.entropy |
| mass | phys.mass |
| metallicity | phys.abund.Z |
| model | *(Model)* |

# 7 Working examples

## 7.1 *The Hyperfine Structure of N2H$^+$*

This example refers to the measurement of the hyperfine structure of the J=1→0 transition in diazenlyium (N$_2$H$^+$) at 93 Ghz (Caselli et al. 1995) toward the cold (kinetic temperature T$_K$~10 K) dense core of the interstellar cloud L1512. Due to the closed-shell $^1\Sigma$ configuration of this molecule, the dominant hyperfine interactions are those between the molecular electric field gradient and the electric quadrupole moments of the two nitrogen nuclei. Together they produce a splitting of the J=1→0 in seven components. The astronomical measurements are much more accurate than those obtainable on the Earth, due to the excellent spectral resolution (~0.18 km s$^{-1}$ FWHM), which correspond to the thermal width at ~20K, much a lower temperature than achievable in the laboratory.

*Table 1 – Observed properties of the N$_2$H$^+$ hyperfine structure components*

| J F$_1$ F → J'F'$_1$F' | ν (MHz) | $\sigma_\nu \Delta\nu$ (MHz) |
|---|---|---|
| 1 0 1 → 0 1 2 | 93176.2650 | 0.0011 |
| 1 2 1 → 0 1 1 | 93173.9666 | 0.0012 |
| 1 2 3 → 0 1 2 | 93173.7767 | 0.0012 |
| 1 2 2 → 0 1 1 | 93173.4796 | 0.0012 |
| 1 1 1 → 0 1 0 | 93172.0533 | 0.0012 |
| 1 1 2 → 0 1 2 | 93171.9168 | 0.0012 |
| 1 1 0 → 0 1 1 | 93171.6210 | 0.0013 |

where $\nu$ is the transition frequency – as derived assuming the same Local Standard Rest velocity for all observed spectral lines – and $\sigma_\nu \Delta\nu$ its relative uncertainty.

Estimates of the N$_2$H$^+$ optical depth, excitation temperature and intrinsic line width were made by fitting the hyperfine splitting complex. They yielded:

- $\tau_{tot} = 7.9 \pm 0.3$
- $T_{ext} = 4.9 \pm 0.1$
- $\Delta v = 183 \pm 1$ m s$^{-1}$

However, the same paper reports evidence for deviations from a single temperature excitation in the following transitions: (F$_1$,F) = (1,2) → (1,2) and (1,0) → (1,1)

We show below an example of instantiation of the current Line Data Model for one of the components of the $N_2H^+$ hyperfine transition (*e.g.* the transition in the first row of *Tab.1*).

In what follows, SI units are assumed whenever pertinent and PhysicalQuantity.error indicates the statistical uncertainty on a measured quantity.

### 7.1.1 The values in the model

In what follows we give the values attached to each of the model items pertinent for the case. For sake of simplicity, we report here the transition in the first row of Table 1 only. Likewise, the class attributes have been given values in pseudo-code way.

**Initial Level** (one QuantumState defined by three QuantumNumber(s)):

- `Line.initialLevel.quantumState.quantumNumber.label := "J"`
- `Line.initialLevel.quantumState.quantumNumber.type := "totalAngularMomentumJ"`
- `Line.initialLevel.quantumState.quantumNumber.description := "Pure quantum number"`
- `Line.initialLevel.quantumState.quantumNumber.numeratorValue := 1`
- `Line.initialLevel.quantumState.quantumNumber.denominatorValue :=1`

<br>

- `Line.initialLevel.quantumState.quantumNumber. label:= "F`$_1$`"`
- `Line.initialLevel.quantumState.quantumNumber.type := "totalAngularMomentumF"`
- `Line.initialLevel.quantumState.quantumNumber.description:= "Resulting angular momentum including nuclear spin for one nucleus; coupling of J and I`$_1$`"`
- `Line.initialLevel.quantumState.quantumNumber.numeratorValue := 0`

<br>

- `Line.initialLevel.quantumState.quantumNumber. label:= "F"`
- `Line.initialLevel.quantumState.quantumNumber.type := "totalAngularMomentumF"`
- `Line.initialLevel.quantumState.quantumNumber.description := "Resulting total angular momentum; coupling of I`$_2$` and F`$_1$`"`
- `Line.initialLevel.quantumState.quantumNumber.numeratorValue := 1`
- `Line.initialLevel.quantumState.quantumNumber.denominatorValue :=1`

**Final Level** (one QuantumState defined by three QuantumNumber(s)):

- `Line.finalLevel.quantumState.quantumNumber. label:= "J"`

- `Line.finalLevel.quantumState.quantumNumber.type := "jtotalAngularMomentum"`
- `Line.finalLevel.quantumState.quantumNumber.description := "Total angular momentum excluding nuclear spins. Pure quantum number"`
- `Line.finalLevel.quantumState.quantumNumber.numeratorValue := 1`
- `Line.initialLevel.quantumState.quantumNumber.denominatorValue :=1`

<br>

- `Line.initialLevel.quantumState.quantumNumber. label:= "F_1"`
- `Line.initialLevel.quantumState.quantumNumber.type := "totalAngularMomentumF"`
- `Line.initialLevel.quantumState.quantumNumber.description:= "Resulting angular momentum including nuclear spin for one nucleus; coupling of J and I_1"`
- `Line.initialLevel.quantumState.quantumNumber.numeratorValue := 0`

<br>

- `Line.initialLevel.quantumState.quantumNumber. label:= "F"`
- `Line.initialLevel.quantumState.quantumNumber.type := "totalAngularMomentumF"`
- `Line.initialLevel.quantumState.quantumNumber.description := "Resulting total angular momentum; coupling of I and J"`
- `Line.initialLevel.quantumState.quantumNumber.numeratorValue := 2`
- `Line.initialLevel.quantumState.quantumNumber.denominatorValue := 1`

**Line** specific attributes:
- `Line.airWavelength.value:= 3.21755760x10^-3`
- `Line.airWavelength.unit.expression:= "m"`
- `Line.airWavelength.unit.scaleSI:= 1`
- `Line.airWavelength.unit.dimEquation:= "L"`

**Process** specific attributes (Broadening):

- `Line.process.type := "Broadening"`
- `Line.process.name := "Intrinsic line width"`
- `Line.observedBroadeningCoefficient.value := 183`
- `Line.observedBroadeningCoefficient.unit.expression := "m/s"`
- `Line.observedBroadeningCoefficient.unit.scaleSI := 1`
- `Line.observedBroadeningCoefficient.unit.dimEquation := "LT-1"`

**Environment** specific attributes:

- `Line.process.model.excitationTemperature.value := 4.9`
- `Line.process.model.excitationTemperature.error := 0.1`
- `Line.process.model.excitationTemperature.unit.expression := "K"`
- `Line.process.model.excitationTemperature.unit.scaleSI := 1`
- `Line.process.model.excitationTemperature.Unit.dimEquation := "K"`
- `Line.process.model.opticalDepth.value := 7.9`
- `Line.process.model.opticalDepth.unit.expression := ""`
- `Line.process.model.opticalDepth.unit.scaleSI := 1`
- `Line.process.model.opticalDepth.unit.dimEquation := ""`
- `Line.process.model.opticalDepth.error := 0.3`

Initial and final (identical) **Specie(s)**:

- `Line.initialElement.name := "N`$_2$`H`$^+$`"`
- `Line.finalElement.name := "N`$_2$`H`$^+$`"`

**Source** specific attributes**:**

- `Line.source.name := "L1512"`

### 7.1.2 JSON representation

```
{
    "Line": {

        "source": {
            "name": "L1512"
        }

        "initialElement": {
            "name": "N2H+"
        }

        "finalElement": {
            "name": "N2H+"
        }

        "initialLevel": {
            "quantumNumber": {
                "label": "J"
                "type": "totalAngularMomentumJ"
                "description": "Pure quantum number"
                "numeratorValue": "1"
                "denominatorValue": "1"
            }
```

```
            "quantumNumber": {
                    "label": "F1"
                    "type": "totalAngularMomentumF"
                    "description": "Resulting angular momentum
including nuclear spin for one nucleus; coupling of J and I1"
                    "numeratorValue": "0"
            }
            "quantumNumber": {
                    "label": "F"
                    "type": "totalAngularMomentumF"
                    "description": 'Resulting total angular
momentum; coupling of I2 and F1"
                    "numeratorValue": "1"
                    "denominatorValue": "1"
            }
        }

        "finalLevel": {
                "quantumNumber": {
                        "label": "J"
                        "type": "jtotalAngularMomentum"
                        "description":  "Total angular momentum
excluding nuclear spins. Pure quantum number"
                        "numeratorValue": "1"
                        "denominatorValue": "1"
                }
                "quantumNumber": {
                        "label": "F1"
                        "type": "totalAngularMomentumF"
                        "description": "Resulting angular momentum
including nuclear spin for one nucleus; coupling of J and I1"
                        "numeratorValue": "0"
                }
                "quantumNumber": {
                        "label": "F"
                        "type": "totalAngularMomentumF"
                        "description": "Resulting total angular
momentum; coupling of I and J"
                        "numeratorValue": "2"
                        "denominatorValue": "1"
                }
        }

        "airWavelength": {
                "value":  "3.21755760x10-3"
                "unit": {
                        "expression": "m"
                        "scaleSI": "1"
                        "dimEquation": "L"
                }
        }
        "process": {
                "type": "Broadening"
                "name": "Intrinsic line width"

                "model": {
                        "excitationTemperature": {
                                "value": "4.9"
                                "error": "0.1"
                                "unit": {
```

```
                                "expression": "K"
                                "scaleSI": "1"
                                "dimEquation": "K"
                        }
                }
                "opticalDepth": {
                        "value": "7.9"
                        "error": "0.3"
                        "unit": {
                                "expression": ""
                                "scaleSI": "1"
                                "dimEquation": ""
                        }
                }
            }
        }

        "observedBroadeningCoefficient": {
                "value": "183"
                "unit" : {
                        "expression": "m/s"
                        "scaleSI": "1"
                        "dimEquation": "LT-1"
                }
        }
    }
}
```

### 7.1.3 UML instantiation diagram

Please note that some physical quantities (marked with an asterisk) have not been fully instanced to simplify the graphics.

## 7.2 Radiative Recombination Continua: a diagnostic tool for X-Ray spectra of AGN

The advent of a new generation of large X-ray observatories is allowing us to obtain spectra of unprecedented quality and resolution on a sizeable number of Active Galactic Nuclei (AGN). This has revived the need for diagnostic tools, which can properly characterize the properties of astrophysical plasmas encompassing the nuclear region, where the gas energy budget is most likely dominated by the high-energy AGN output.

Among these spectra diagnostics, Radiative Recombination Continua (RRC) play a key role, as they unambiguously identify photoionized plasmas, and provide unique information on their physical properties. The first quantitative studies which recognized the importance of RRC in X-ray spectra date back to the early '90, using *Einstein* (Liedahl et al. 1991; Kahn & Liedahl 1991) and ASCA (Angelini et al. 1995) observations. The pioneer application of the RRC diagnostic to AGN is due to Kinkhabwala et al. (2002; K02), who analysed a long XMM-Newton/RGS observation of the nearby Seyfert 2 galaxy NGC1068 (z=0.003793, corresponding to a recession velocity of 1137 km s$^{-1}$). We will refer to the results reported in their paper hereafter.

K02 report the detection of RRC from 6 different ionic species. Their observational properties are shown in Tab.3. The RRC temperature $kT_e$ is derived from the RRC profile fit, as the width of the RRC profile $\Delta E \approx kT_e$. The average RRC photon energy is $E \approx I + kT_e$, where *I* is the ionization potential of the recombined state. If the plasma is highly over ionised ($kT \ll I$) – as expected in X-ray photoionized nebulae (Kallman & McCray 1982) – then $\Delta E/E \approx kT_e/I$. Therefore, the specification of $kT_e$ (extracted from Tab.2 in K02) and $I$ (extracted from table of photo ionisation potentials) is enough to know the energy of the feature.

*Tab.3 – Properties of the RRC features in the XMM-Newton/RGS spectrum of NGC1068*

| Ion | $kT_e$ (eV) | Flux (10$^{-4}$ ph cm$^{-2}$ s$^{-1}$) | I (eV) |
|---|---|---|---|
| CV | 2.5±0.5 | 4.3±0.4 | 392.1 |
| CVI | 4.0±1.0 | 2.8±0.3 | 490.0 |
| NVI | 3.5±2.0 | 2.1±0.2 | 552.1 |
| NVII | 5.0±3.0 | 1.1±0.1 | 667.1 |
| OVII | 4.0±1.3 | 2.4±0.2 | 739.3 |
| OVIII | 7.0±3.5 | 1.2±0.1 | 871.4 |

### 7.2.1 The values in the model

**Initial Level** description:
- `Line.initialLevel.quantumState.quantumNumber.label:= "n"`
- `Line.initialLevel.quantumState.quantumNumber.type:= "nPrincipal"`
- `Line.initialLevel.quantumState.quantumNumber.numeratorValue:=1`
- `Line.initialLevel.quantumState.quantumNumber.denominatorValue:= 1`

**Final Level**:
- `Line.finalLevel.quantumState.quantumNumber.label:= "n"`
- `Line.finalLevel.quantumState.quantumNumber.type:= "nPrincipal"`
- `Line.finalLevel.quantumState.quantumNumber.numeratorValue:= 1`
- `Line.finalLevel.quantumState.quantumNumber.denominatorValue:= 1`

**Initial Element**:
- `Line.initialElement.species.name := "CVI"`

**Final Element**:
- `Line.finalElement.species.name := "CV"`

(Observed) **Line** specific attributes

- `Line.wavelength.value := 394.6`
- `Line.wavelength.unit.expression := "eV"`
- `Line.wavelength.Unit.scaleSI := 1.6E-19`
- `Line.wavelength.Unit.dimEquation := "ML2T-2"`

- `Line.observedFlux.value := 2.8E-4`
- `Line.observedFlux.error := 0.3E-4`
- `Line.observedFlux.unit.expression := "photons*cm-2*s-1"`
- `Line.observedFlux.unit.scaleSI = 1.E4`
- `Line.observedFlux.unit.dimEquation := "L-2T-1"`

- `Line.transitionType := "Radiative Recombination Continuum"`

**Process** specific attributes
- `Line.Process.type := "Energy shift"`
- `Line.Process.name := "Cosmological redshift"`

- `Line.observedShiftingCoefficient.value := 1137`
- `Line.observedShiftingCoefficient.unit.expression := "km/s"`
- `Line.observedShiftingCoefficient.unit.scaleSI := 1.E3`
- `Line.observedShiftingCoefficient.unit.dimEquation := "MT-1"`

**Environment** specific attributes:

- `Line.environment.temperature.value := 1.9E5`
- `Line.environment.temperature.unit.expression := "K"`
- `Line.environment.temperature.unit.scaleSI := 1`
- `Line.environment.temperature.unit.dimEquation := "K"`

**Source** specific attributes**:**

- `Line.source.name := "NGC1068"`

### 7.2.2 JSON representation

```
{
    "Line": {
        "transitionType":  "Radiative Recombination Continuum"
        "wavelength": {
            "value":  "394.6"
            "unit": {
                "expression": eV"
                "scaleSI": "1.6E-19"
                "dimEquation": "ML2T-2"
            }
        }
        "observedFlux": {
            "value":  "2.8E-4"
            "error": "0.3E-4"
            "unit": {
                "expression": photons*cm-2*s-1"
                "scaleSI": "1.E4"
                "dimEquation": "L-2T-1"
            }
        }
        "source": {
            "name": "NGC1068"
        }
        "initialElement": {
            "name": "CVI"
        }
        "finalElement": {
            "name": "CV"
        }
        "initialLevel": {
            "quantumNumber": {
                "label": "n"
                "type": "nPrincipal"
```

```
                        "numeratorValue": "1"
                        "denominatorValue": "1"
                    }
                }
                "finalLevel": {
                    "quantumNumber": {
                        "label": "n"
                        "type": "nPrincipal"
                        "numeratorValue": "1"
                        "denominatorValue": "1"
                    }
                }
                "process": {
                    "type": "Energy shift"
                    "name": "Cosmological redshift"
                }
                "observedShiftingCoefficient": {
                    "value": "1137"
                    "unit" : {
                        "expression": "km/s"
                        "scaleSI": "1.E3"
                        "dimEquation": "MT-1"
                    }
                }
                "environment": {
                    "temperature": {
                        "value": "1.9E5"
                        "unit": {
                            "expression": "K"
                            "scaleSI": "1"
                            "dimEquation": "K"
                        }
                    }
                }
            }
        }
```

### 7.2.3 UML Instantiation diagram

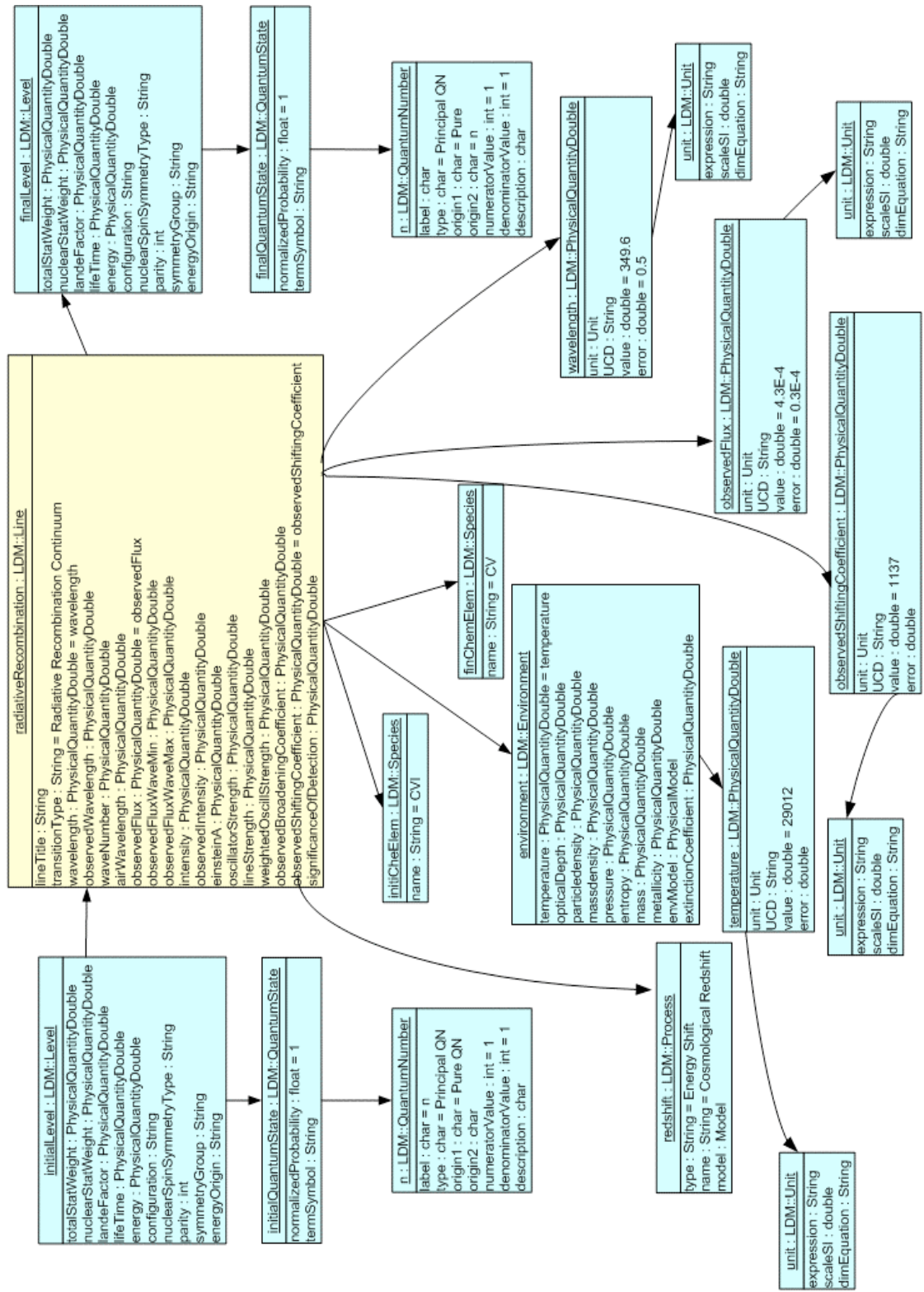

# 8   Appendix A: List of Atomic Elements

List of Elements extracted from the **IUPAC Commission on Atomic Weights and Isotopic Abundances**. (http://www.chem.qmul.ac.uk/iupac/)

**List of Elements in Atomic Number Order.**

| At No | Symbol | Name | Notes |
|---|---|---|---|
| 1 | H | Hydrogen | 1, 2, 3 |
| 2 | He | Helium | 1, 2 |
| 3 | Li | Lithium | 1, 2, 3, 4 |
| 4 | Be | Beryllium | |
| 5 | B | Boron | 1, 2, 3 |
| 6 | C | Carbon | 1, 2 |
| 7 | N | Nitrogen | 1, 2 |
| 8 | O | Oxygen | 1, 2 |
| 9 | F | Fluorine | |
| 10 | Ne | Neon | 1, 3 |
| 11 | Na | Sodium | |
| 12 | Mg | Magnesium | |
| 13 | Al | Aluminum | |
| 14 | Si | Silicon | 2 |
| 15 | P | Phosphorus | |
| 16 | S | Sulfur | 1, 2 |
| 17 | Cl | Chlorine | 3 |
| 18 | Ar | Argon | 1, 2 |
| 19 | K | Potassium | 1 |
| 20 | Ca | Calcium | 1 |
| 21 | Sc | Scandium | |
| 22 | Ti | Titanium | |
| 23 | V | Vanadium | |
| 24 | Cr | Chromium | |
| 25 | Mn | Manganese | |
| 26 | Fe | Iron | |
| 27 | Co | Cobalt | |
| 28 | Ni | Nickel | |
| 29 | Cu | Copper | 2 |

| 30 | Zn | Zinc | |
| 31 | Ga | Gallium | |
| 32 | Ge | Germanium | |
| 33 | As | Arsenic | |
| 34 | Se | Selenium | |
| 35 | Br | Bromine | |
| 36 | Kr | Krypton | 1, 3 |
| 37 | Rb | Rubidium | 1 |
| 38 | Sr | Strontium | 1, 2 |
| 39 | Y | Yttrium | |
| 40 | Zr | Zirconium | 1 |
| 41 | Nb | Niobium | |
| 42 | Mo | Molybdenum | 1 |
| 43 | Tc | Technetium | 5 |
| 44 | Ru | Ruthenium | 1 |
| 45 | Rh | Rhodium | |
| 46 | Pd | Palladium | 1 |
| 47 | Ag | Silver | 1 |
| 48 | Cd | Cadmium | 1 |
| 49 | In | Indium | |
| 50 | Sn | Tin | 1 |
| 51 | Sb | Antimony | 1 |
| 52 | Te | Tellurium | 1 |
| 53 | I | Iodine | |
| 54 | Xe | Xenon | 1, 3 |
| 55 | Cs | Caesium | |
| 56 | Ba | Barium | |
| 57 | La | Lanthanum | 1 |
| 58 | Ce | Cerium | 1 |
| 59 | Pr | Praseodymium | |
| 60 | Nd | Neodymium | 1 |
| 61 | Pm | Promethium | 5 |
| 62 | Sm | Samarium | 1 |
| 63 | Eu | Europium | 1 |
| 64 | Gd | Gadolinium | 1 |
| 65 | Tb | Terbium | |
| 66 | Dy | Dysprosium | 1 |
| 67 | Ho | Holmium | |
| 68 | Er | Erbium | 1 |
| 69 | Tm | Thulium | |
| 70 | Yb | Ytterbium | 1 |
| 71 | Lu | Lutetium | 1 |
| 72 | Hf | Hafnium | |
| 73 | Ta | Tantalum | |
| 74 | W | Tungsten | |
| 75 | Re | Rhenium | |
| 76 | Os | Osmium | 1 |
| 77 | Ir | Iridium | |
| 78 | Pt | Platinum | |
| 79 | Au | Gold | |
| 80 | Hg | Mercury | |
| 81 | Tl | Thallium | |
| 82 | Pb | Lead | 1, 2 |
| 83 | Bi | Bismuth | |
| 84 | Po | Polonium | 5 |
| 85 | At | Astatine | 5 |
| 86 | Rn | Radon | 5 |
| 87 | Fr | Francium | 5 |
| 88 | Ra | Radium | 5 |
| 89 | Ac | Actinium | 5 |
| 90 | Th | Thorium | 1, 5 |
| 91 | Pa | Protactinium | 5 |
| 92 | U | Uranium | 1, 3, 5 |
| 93 | Np | Neptunium | 5 |
| 94 | Pu | Plutonium | 5 |
| 95 | Am | Americium | 5 |
| 96 | Cm | Curium | 5 |
| 97 | Bk | Berkelium | 5 |
| 98 | Cf | Californium | 5 |
| 99 | Es | Einsteinium | 5 |
| 100 | Fm | Fermium | 5 |
| 101 | Md | Mendelevium | 5 |
| 102 | No | Nobelium | 5 |
| 103 | Lr | Lawrencium | 5 |
| 104 | Rf | Rutherfordium | 5, 6 |
| 105 | Db | Dubnium | 5, 6 |
| 106 | Sg | Seaborgium | 5, 6 |
| 107 | Bh | Bohrium | 5, 6 |

| 108 | Hs  | Hassium      | 5, 6           |
| 109 | Mt  | Meitnerium   | 5, 6           |
| 110 | Ds  | Darmstadtium | 5, 6           |
| 111 | Rg  | Roentgenium  | 5, 6           |
| 112 | Uub | Ununbium     | 5, 6           |
| 114 | Uuq | Ununquadium  | 5, 6           |
| 116 | Uuh | Ununhexium   | see Note above |
| 118 | Uuo | Ununoctium   | see Note above |

1. Geological specimens are known in which the element has an isotopic composition outside the limits for normal material. The difference between the atomic weight of the element in such specimens and that given in the Table may exceed the stated uncertainty.

2. Range in isotopic composition of normal terrestrial material prevents a more precise value being given; the tabulated value should be applicable to any normal material.

3. Modified isotopic compositions may be found in commercially available material because it has been subject to an undisclosed or inadvertent isotopic fractionation. Substantial deviations in atomic weight of the element from that given in the Table can occur.

4. Commercially available Li materials have atomic weights that range between 6.939 and 6.996; if a more accurate value is required, it must be determined for the specific material [range quoted for 1995 table 6.94 and 6.99].

5. Element has no stable nuclides. The value enclosed in brackets, e.g. [209], indicates the mass number of the longest-lived isotope of the element. However three such elements (Th, Pa, and U) do have a characteristic terrestrial isotopic composition, and for these an atomic weight is tabulated.

The names and symbols for elements 112-118 are under review. The temporary system recommended by J Chatt, *Pure Appl. Chem.*, **51**, 381-384 (1979) is used above. The names of elements 101-109 were agreed in 1997 (See *Pure Appl. Chem.,* 1997, **69**, 2471-2473),for element 110 in 2003 (see *Pure Appl. Chem., 2003, 75, 1613-1615*) and for element 111 in 2004 (see *Pure Appl. Chem., 2004, 76, 2101-2103*).

# 9 APPENDIX B: Description of couplings for atomic Physics

## 9.1 LS coupling

Usually the strongest interactions among the electrons of an atom are their mutual Coulomb repulsions. These repulsions affect only the orbital angular momenta, and not the spins. It is thus most appropriate to first couple together all the orbital angular momenta to give eigenfunctions
of $\mathbf{L}^2$ and $\mathbf{L}_Z$, with $\mathbf{L}$ the total orbital angular momentum of the atom. Similarly all spins are coupled together to give the eigenfunctions of $\mathbf{S}^2$ and $\mathbf{S}_Z$, with $\mathbf{S}$ the total spin angular momentum; then $\mathbf{L}$ and $\mathbf{S}$ are coupled together to give eigenfunctions of $\mathbf{J}^2$ and $\mathbf{J}_Z$, where $\mathbf{J}=\mathbf{L}+\mathbf{S}$.

When the coupling conditions within an atom correspond closely to pure LS-coupling conditions, then the quantum states of an atom can be accurately described in terms of LS-coupling quantum numbers:

Giving values of L and S specifies a term, or more precisely a ``LS term'', because on may also refer to ``terms'' of a different sort when discussing other coupling schemes (In order to completely specify a term it is necessary to give not only values of L and S, but also values of all lower-order quantum numbers, such as $n_i l_i$.

- Giving values of L, S, J specifies a level
- Giving values of L, S, J, $M_J$ specifies a state
- The value of (2S+1) is called the multiplicity of the term

For LS-coupled functions, the notation introduced by Russel and Saunders is universally used : $^{2S+1}L_J$, where numerical values are to be substituted for (2S+1) and J, and the appropriate letter symbol is used for L (S, P, ..); except when discussing the Stark or Zeeman effect, there is usually no need to specify the value of $M_J$.

## 9.2 jj coupling

With increasing Z, the spin-orbit interactions become increasingly more important; in the limit in which these interactions become much stronger than the Coulomb terms, the coupling conditions approach pure jj coupling.

In the jj-coupling scheme, basis functions are formed by first coupling the spin of each electron to its own orbital angular momentum, and then coupling together the various resultants $j_i$ in some arbitrary order to obtain the total angular momentum **J**. For two-electron configurations, the coupling scheme may be described by the condensed notation $[(l_1 s_1)j_1, (l_2, s_2)j_2]JM_J$ with the usual jj-notation for energy levels $(j_1, j_2)_J$ [analogous to the Russel Saunders notation $^{2S+1}L_J$].

## 9.3 jK coupling

For configurations containing only two electron outside of closed shells, the common limiting type of pair coupling (energy levels tend to appear in pairs), jK coupling, occurs when the strongest interaction is the spin-orbit interaction of the more tightly bound electron, and the next strongest interaction is the spin-independent (direct) portion of the Coulomb interaction between the 2 electrons.

The corresponding angular-momentum coupling scheme is $l_1 + s_1 = j_1$, $j_1 + l_2 = K$, $K + s_2 = J$, or notation $\{[(l_1 s_1)j_1, l_2]K, s_2\}JM$ with the standard energy level notation $j_1[K]_J$.

## 9.4 LK coupling

The other limiting form of pair coupling is called LK (or Ls) coupling. In two-electron configurations, it corresponds to the case in which the direct Coulomb interaction is greater then the spin-orbit interaction of either electron, and the spin-orbit interaction of the inner electron is next most important. The coupling scheme is $l_1 + l_2 = L$, $L + s_1 = K$, $K + s_2 = J$, or notation $\{[(l_1 l_2)L, s_1]K, s_2\}JM$ with the standard energy level notation $L[K]_J$.

## 9.5 Intermediate coupling

Frequently the coupling conditions do not lie particularly close even to one of these four cases; such situation is referred to as intermediate coupling. The energy levels can only be labelled in terms of the least objectionable of the four pure-coupling schemes (with the understanding that these labels may give a poor description of the true angular-momentum properties of the corresponding quantum states). In many cases, however, the coupling conditions are so hopelessly far from any pure-coupling scheme that it is meaningless to do anything more than label the energy levels and quantum states by means of serial numbers or some similar arbitrary device, or to list the values of the largest few eigenvector components (or the squares thereof) in the expansion of the total wavefunction.

"The wavefunctions of levels are often expressed as eigenvectors that are linear combinations of basis states in one of the standard coupling schemes. Thus, the wave function $\Psi(\alpha J)$ for a level labeled $\alpha J$ might be expressed in terms of normalized *LS* coupling basis states $\Phi(\gamma SLJ)$: $\Psi(\alpha J)=\sum_{\gamma SL} c(\gamma SLJ)\Phi(SLJ)$ The $c(\gamma SLJ)$ are expansion coefficients, and $\sum_{\gamma SL} |c(gSLJ)|^2 = 1$ (Martin & Wiese)

The expansion coefficients are called "mixingCoefficient" in this document.

The squared expansion coefficients for the various $\gamma SL$ terms in the composition of the $\alpha J$ level are conveniently expressed as percentages, whose sum is 100%. The notation for RS basis states has been used only for concreteness; the eigenvectors may be expressed in any coupling scheme, and the coupling schemes may be different for different configurations included in a single calculation (with configuration interaction). « Intermediate coupling » conditions for a configuration are such that calculations in both LS and jj coupling yield some eigenvectors representing significant mixtures of basis states.

The largest percentage in the composition of a level is called the *purity* of the level in that coupling scheme. The coupling scheme (or combination of coupling schemes if more than one configuration is involved) that results in the largest average purity for all the levels in a calculation is usually best for naming the levels. With regard to any particular calculation, one does well to remember that, as with other calculated quantities, the resulting eigenvectors depend on a specific theoretical model and are subject to the inaccuracies of whatever approximations the model involves. Theoretical calculations of experimental energy level structures have yielded many eigenvectors having significantly less than 50% purity in any coupling scheme. Since many of the corresponding levels have nevertheless been assigned names by spectroscopists, some caution is advisable in the acceptance of level designations found in the literature. »